\documentclass[aps,prc,onecolumn,showpacs,amsmath,amssymb,nofootinbib,11pt]{revtex4-2}
\usepackage{graphicx}
\usepackage{color}
\usepackage{times}
\usepackage{inputenc}
\usepackage{bm}
\usepackage{ulem}
\usepackage{multirow}
\usepackage{float}
\usepackage{url}
\usepackage{natbib}
\usepackage{mathrsfs}
\usepackage{physics}
\usepackage{comment}
\usepackage[table,xcdraw]{xcolor}
\usepackage{booktabs,makecell}
\usepackage{comment}
\usepackage[colorlinks=true,citecolor=blue,urlcolor=blue,linkcolor=blue]{hyperref}

\begin{document}
\title{Decoding Dark Matter Admixed Neutron Stars: From Static Structure to Rotational Deformation}
\author{Pinku Routaray$^{1}$}
\email{routaraypinku@gmail.com}
\author{Abirbhav Chakrawarty$^{1}$}
\author{N. K. Patra$^2$}
\author{Bharat Kumar$^{1}$}

\affiliation{\it $^{1}$ Department of Physics and Astronomy, National Institute of Technology, Rourkela 769008, India}
\affiliation{\it $^2$ School of Science and Engineering, The Chinese University of Hong Kong, Shenzhen (CUHKShenzhen), Guangdong, 518172, China}
\date{\today}

\begin{abstract}
In this study, we investigate the impacts of dark matter (DM) on the properties of both static and rotating neutron stars utilizing a self-interacting DM model, motivated by the neutron decay anomaly. DM-admixed NSs are modeled by assuming chemical equilibrium between ordinary matter and the dark sector, treating a single-fluid Tolman-Oppenheimer-Volkoff (TOV) framework. By treating the DM interaction strength ($G$) as a free parameter, we explore its influence on NS properties, considering a broad range of equations of state (EoSs). Using the mass-radius constraints from NICER pulsar measurements, we constrain the DM interaction strength for each EoS via a likelihood analysis. Extending this model to rotating NSs, we analyze how centrifugal forces associated with increasing angular velocity ($\Omega$) enhance both mass and radius, causing deformation. We assess the impact of DM on rotational deformation by calculating the eccentricity, highlighting the interplay between DM and rotational forces. Since both DM and rotation simultaneously influence NS properties, we compute the relative changes in mass and radius across varying $G$ and $\Omega$ values to quantify their combined effects.
\end{abstract}
\maketitle

\section{Introduction}
Neutron stars (NSs) are ideal for studying physics in extreme conditions, where high densities and strong gravitational fields push matter beyond nuclear saturation density ($n_0 = 0.15 \, \text{fm}^{-3}$) \cite{Lattimer_2001, Lattimer_Rev_2012, Lattimer_Rev_2021}. Understanding and validating the EOSs at these densities is crucial, as the behavior of matter in this regime remains poorly understood. Key properties of NSs, such as mass, radius, compactness, and rotational deformation, are directly influenced by the EOS and can be theoretically modeled to align with observational data \cite{Abbott_2017, Abbott_2018, Tuhin_2018, Miller_2019, Riley_2019, Miller_2021, Riley_2021}. While the TOV equations provide an accurate framework for modeling non-rotating, spherically symmetric NSs \cite{Tolman_1939, Oppenheimer_1939}, rotation introduces significant deformations that necessitate the use of more advanced approaches, such as the Hartle-Thorne formalism for slowly rotating stars \cite{Hartle_1967, Hartle-Thorne_1968}. Both of these approaches show significant impact for the DM-admixed EOS \cite{Goldman_1989, Narain06, Kouvaris_2008, Kouvaris_2010, Grigorious_2017, Nelson_2019, Davood_2022, Glendenning_rns_1992, Ishfaq_2021, Cronin_rns-DM_2023, Konstantinou_rns-DM_2024,pinku_tf-mf_2025}.

DM is believed to constitute a substantial portion of the Universe's mass-energy content, with its existence supported by various astrophysical and cosmological observations, including cosmic microwave background radiation, large-scale structure formation, and galaxy rotation curves \cite{bertone2005particle}. Numerous DM candidates have been proposed so far \cite{Goldman_1989, bertone2005particle, KHLOPOV_2013, bauer2019yet, Yohei-Robert_2024}. In the context of NSs, different models have been developed to study the effects of DM. Some studies employ a two-fluid framework, exploring gravitational interactions that suggest the possible presence of DM within the NS or extending to a halo structure \cite{Sandin_2009, Nelson_2019, arpan_two-fluid_2022, Davood_2022, Giangrandi_2023,Zeinab-Fuzzy-DM_2023}. Alternatively, other studies explore non-gravitational interactions by treating NS and DM as a single fluid, considering DM presence within the NS \cite{Goldman_1989, Kouvaris_2008, Kouvaris_2010, Kouvaris_2011, Ciarcelluti_2011, Grigorious_2017, harishmnras_2020, harishprd_2021, pinku_prd_2023, pinku_mnras_2023,pinku_ijmpe_2024}. Despite ongoing research, the precise nature and properties of DM remain elusive, continuing to fuel scientific inquiry.

The EOS for cold, dense nuclear matter remains uncertain, which unveils a problem and adds further complexity when performing the calculation of DM-admixed NS characteristics. This uncertainty implies that the mass-radius relationship of DM-admixed NSs depends heavily on the choice of EOS used to describe the baryonic matter alongside DM. Observations from the NICER X-ray telescope, which measures pulsed X-ray emissions from millisecond rotation-powered pulsars, provide a method to determine both mass and radius \cite{Gendreau_2012}. However, conventional NICER data analysis assumes that pulsars lack any DM contribution. Addressing this limitation, Rutherford {\it et al.} \cite{Rutherford_2023} proposed an analytical approach that relies on NICER measurements to establish simultaneous limits on both baryonic and DM EOSs. Their study specifically considers scenarios where all DM is confined within the baryonic surface of the NS, forming a DM core. Following to this work, Andreas {\it et al.} \cite{Konstantinou_rns-DM_2024} analyzed NICER data using a two-fluid model for rotating DM-admixed NSs, where DM interacts with the NS solely through gravity. Additionally, several studies have explored the properties of rotating NSs using various frameworks \cite{Glendenning_rns_1992, Chubarian_1999, Dhiman_2007, Most_2020, Ishfaq_2021, Guha_2021, Pattersons-RNS_2021, Silva_2023, John_RNS-DM_2023, Lopes-RNS_2024, Konstantinou_rns-DM_2024}. Building on these works, it becomes imperative to investigate the interplay of DM and rotation in NS properties under the single-fluid approximation, which accounts for non-gravitational interactions between DM and baryonic matter, while assuming that DM is confined within the NS.

In this work, we investigate the effects of DM on both static and rotating NSs using a self-interacting DM model inspired by the neutron decay anomaly \cite{Fornal_anomaly_dark_2018, Fornal_2019, Fornal_nucl-stability_2020, Motta_anomaly_2018, Husain_2022, Swarnim_2023}. This anomaly, arising from inconsistencies in neutron lifetime measurements between beam and bottle experiments, suggests the existence of a decay channel involving DM. Under this framework, DM accumulates in the NS core, significantly influencing its structural and dynamic properties. By assuming chemical equilibrium between DM and baryonic matter, the system can be effectively modeled using a single-fluid approximation \cite{Husain_2022, Swarnim_2023}. To capture these effects, we develop a DM-admixed NS EOS within the relativistic mean-field (RMF) framework, which includes mesonic interactions to describe baryonic matter \cite{serot1992relativistic, serot1997recent}. A range of EOSs is considered, which is able to satisfy chiral effective field theory ($\chi$-EFT) constraints at low density for pure neutron matter, and also they can satisfy multi-messenger observations. After achieving the DM-admixed NS, we investigate how the variation of DM interaction strength affects the properties of static NS for each EOS. Additionally, we calculate the DM fraction during the TOV integration and examine its relationship with $G$ and EOS stiffness. We constrain the DM interaction strength within the DM-admixed NS using a likelihood analysis. This analysis is performed by incorporating mass-radius measurements from NICER X-ray observations of the pulsars PSR J0030+0451 \cite{Miller_2019, Riley_2019} and PSR J0740+6620 \cite{Miller_2021, Riley_2021}. Further, we extend our study for DM-admixed rotating NS by employing the Hartle-Thorne formalism for slow rotation. The rotational effects are considered by varying the angular velocity for a rotating neutron star (RNS), and its value is chosen considering constraints of the fastest rotating pulsar PSR J1748-2446ad, which spins at angular velocity $\Omega=4498 \ {\rm s}^{-1}$ \cite{hessels2006radio}. Then, for DM-admixed rotating NS, we systematically explore the variations in DM interaction strength and angular velocity, quantifying deviations in rotational mass and radius compared to a static, DM-free NS. Additionally, for the rotating NS case, we perform a likelihood analysis to examine how rotation influences the DM interaction strength in light of observational constraints. Finally, we investigate how the presence of DM affects the deformation of the rotating NS.

This paper is structured as follows: Section \ref{subsec:eoshadron} introduces the RMF formalism and considered energy density functional in the current study. Section \ref{subsec:dm_model} discusses details about the DM model. Section \ref{subsec:steq} discusses the single-fluid TOV equations for determining the equilibrium structure of static DM-admixed NS. Section \ref{subsec:rns} outlines the Hartle-Thorne formalism for slowly rotating NS. The results are presented in Section \ref{sec:results}, followed by a summary and conclusions in Section \ref{sec:summary}.

\section{Formalism}
\subsection{Hadronic Model}
\label{subsec:eoshadron}
In the RMF formalism, nucleons interact via the exchange of mesons, and we consider three types of mesons: the scalar $\sigma$, the vector $\omega$, and the isovector $\rho$ mesons. The Lagrangian describing nucleonic matter is given by \cite{FURNSTAHL_1996,Singh_2014,Kumar_2018,pinku_jcap_2023,Probit_2024,Probit_prc_2024}:
\begin{equation}
    \begin{aligned}
        \mathcal{L}_{\text{nuc}} &= \sum_{\alpha=n,p} \bar{\psi}_{\alpha} \left[ \gamma_{\mu} \left( i\partial^{\mu} - g_{\omega} \omega^{\mu} - \frac{1}{2} g_{\rho} {\tau} \cdot {\rho}^{\mu} \right) \right. \\
        & \left. - \left( M_N -\vphantom{\frac{a}{b}} g_{\sigma} \sigma \right) \right] \psi_{\alpha} + \frac{1}{2}\partial_{\mu}\sigma\partial^{\mu}\sigma - \frac{1}{2}m_{\sigma}^2\sigma^2 \\
        & +\frac{\zeta_0}{4!}g_\omega^2(\omega^{\mu}\omega_{\mu})^2-\frac{\kappa_3}{3!}\frac{g_{\sigma}m_{\sigma}^2\sigma^3}{M_N}-\frac{\kappa_4}{4!}\frac{g_{\sigma}^2m_{\sigma}^2\sigma^4}{M_N^2} \\
        & + \frac{1}{2}m_\omega^2\omega_{\mu}\omega^{\mu} - \frac{1}{4}W^{\mu \nu}W_{\mu\nu} + \frac{1}{2}m_\rho^2{\rho^\mu}\cdot{\rho_\mu} \\
        & - \frac{1}{4}{R^{\mu\nu}}\cdot{R_{\mu\nu}} - \Lambda_\omega g_\omega^2 g_\rho^2 (\omega^\mu \omega_\mu)({\rho^\mu}\cdot{\rho_\mu)}
    \end{aligned}
    \label{eq:lagrangian}
\end{equation}

Here, $\psi$ represents the nucleonic Dirac spinor, while $\sigma$, $\omega^\mu$, and $\rho^\mu$ denote the sigma, omega, and rho meson fields, respectively. $M_N$, $m_\sigma$, $m_\omega$, and $m_\rho$ are the masses of the nucleons and mesons. The parameters $g_\sigma$, $g_\omega$, and $g_\rho$ are the meson coupling constants, whereas $\kappa_3$ and $\kappa_4$ are the third- and fourth-order self-coupling constants of the scalar meson field. The constants $\zeta_0$ and $\Lambda_\omega$ represent the vector meson self-coupling and the vector-isovector meson coupling, respectively. $W_{\mu\nu} = \partial_\mu \omega_\nu - \partial_\nu \omega_\mu$ and $R_{\mu\nu} = \partial_\mu \rho_\nu - \partial_\nu \rho_\mu$ are the field tensors. Finally, $\tau$ is the isospin operator.

For the leptonic contribution, we consider two leptons in the system: $e^-$ and $\mu^-$, and the corresponding Lagrangian contribution is given by, 
\begin{equation}
\mathcal{L}_{\text{lep}} = \sum_{l=e, \mu} \bar{\psi}_k (i\gamma^\mu \partial\mu - m_k) \psi_k
\end{equation}

where $\psi_k$ and $m_k$ represent the leptonic Dirac spinor and mass, respectively.

Hence, the energy density of NS can be written as, 
\begin{equation}
    \begin{aligned}
        {\cal E}_{NS} = & \sum_{\alpha = n, p} \frac{1}{8\pi^2} \left[k_\alpha E_\alpha^3 + k_\alpha^3 E_\alpha - M^{\star} \ln \left(\frac{k_\alpha + E_\alpha}{M^\star}\right)\right] \\
        & + \frac{1}{2}m_\sigma ^ 2 \sigma_0^2 + \frac{\kappa_3}{3!} \frac{m_s^2\left(g_\sigma \sigma_0 \right)^3}{g_s^2M_N} + \frac{\kappa_4}{4!} \frac{m_s^2\left(g_\sigma \sigma_0 \right)^4}{g_s^2M_N^2} \\ 
        & - \frac{1}{2}m_\omega^2 \omega_0^2 - \frac{1}{4!}\frac{\zeta_0\left(g_\omega \omega_0\right)^4}{g_\omega^2} - \frac{1}{2}m_\rho^2 \rho_0^2 - \Lambda_\omega\left(g_\rho g_\omega \rho_0 \omega_0\right)^2 \\ 
        & + \sum_{l = e^-, \mu^-}\frac{1}{\pi^2} \int_{0}^{k_i} k^2\sqrt{k^2 + m_i^2} \,dk 
    \end{aligned}
\end{equation}
where $k_\alpha$ and $E_\alpha = \sqrt{k_\alpha^2 + M^{\star^2}}$ are the Fermi momentum and Fermi energy of the $\alpha^{th}$ nucleonic species respectively. Here $M^\star$ represents the effective mass of the nucleon and is given by $M^\star = M_N - g_\sigma \sigma$.

With expression for energy density in hand, the expression for pressure ($P$) can be obtained using Gibbs-Duhem relation
\begin{equation}
    P_{NS} = \sum_{N=\alpha, \ l} \mu_N n_N - {\cal E}_{NS}
\end{equation}

Chemical potential of the $\alpha^{th}$ nucleonic species $\mu_{\alpha}$ is given by,
\begin{equation}
    \mu_{\alpha} = E_{\alpha} + g_\omega \omega_0 + \frac{1}{2}\tau_{3\alpha}\rho_0 
\end{equation}
where $\tau_{3\alpha}$ is the third component of nucleonic isospin operator. 

For leptons, the chemical potential is given by,
\begin{equation}
    \mu_l = \sqrt{k_l^2 + m_l^2}
\end{equation}

The NS is in beta equilibrium and maintains the charge neutrality condition as follows, 
\begin{align*}
    \mu_n &= \mu_p + \mu_e, \\
    \mu_\mu &= \mu_e, \\
    n_p &= n_e + n_\mu.
\end{align*}

For the current study, we employ four energy density functionals named Hornick1, Hornick2, Hornick3 and Hornick4 within the RMF formalism, which has been constructed by varying the effective mass of the nucleon to satisfy the nuclear saturation parameters extracted from nuclear experiments and multi-messenger data, and details are given in ref. \cite{Hornick_2018}. The choice of model parameters is given in the Table.\ref{tab:parameter}.
\begin{table}
\caption{The masses of the $\sigma$, $\omega$, and $\rho$ mesons are fixed at 550 MeV, 783 MeV, and 770 MeV, respectively, for all models. The coupling parameter $\zeta=0$ is set for every model. The nucleon mass ($M_N$) is taken as 939 MeV.}
\renewcommand{\tabcolsep}{0.1cm}
\renewcommand{\arraystretch}{1.3}
\scalebox{1.0}{
\begin{tabular}{cccccccc}
\hline \hline
Model & $M^\star/M_N$ & $g_\sigma$   & $g_\omega$   &$g_\rho$   & $\kappa_3$   &$\kappa_4$ & $\Lambda_{\omega}$\\
\hline
 Hornick1 & 0.55 & 11.552  & 13.566  & 11.871  & 1.5471  &  -6.6419 & 0.0295 \\
 Hornick2 & 0.60 & 10.993 & 12.708 & 10.651 & 1.6729 & -6.6740 & 0.0272 \\
 Hornick3 & 0.65 & 10.429 & 11.774 & 10.186 & 1.9556 & -7.0031 & 0.0278 \\
 Hornick4 & 0.70 & 9.846 & 10.746 & 9.982 & 2.4385 & -7.3696 & 0.0314 \\
\hline \hline
\end{tabular}}
\label{tab:parameter}
\end{table}
\begin{figure}
   \centering
   \includegraphics[width = 0.5\textwidth]{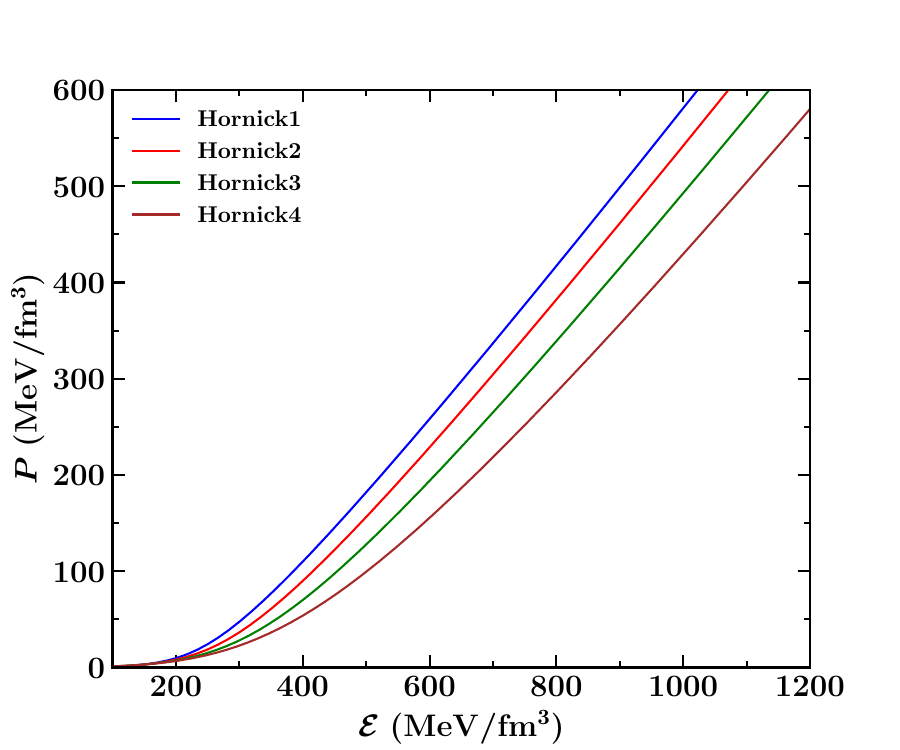}
   \caption{The hadronic EOSs such as Hornick1 \cite{Hornick_2018}, Hornick-2 \cite{Hornick_2018}, Hornick-3 \cite{Hornick_2018}, Hornick-4 \cite{Hornick_2018} is shown.}
    \label{fig:eos_hnk}
\end{figure}
The EOSs for pure hadronic models are shown in the Fig. \ref{fig:eos_hnk}. The broader range of EOSs with varying effective mass of nucleon is considered, which directly influences their stiffness. Specifically, the Hornick1 EOS, characterized by a smaller effective mass of nucleon, is the stiffest among the EOSs presented, while Hornick-4, with a higher effective mass of nucleon as given in ref. \cite{Hornick_2018}, exhibits the softest behavior. The adoption of the BPS crust in the present study ensures a unified low-density behavior across the EOSs \cite{BPS_crust_1971}. Additionally, at lower energy densities, these EOSs remain soft, which aligns them with constraints imposed by $\chi$-EFT calculations for pure neutron matter \cite{Hornick_2018}. This consistency with $\chi$-EFT strengthens the physical reliability of these EOSs at low densities, while their divergent behaviors at higher densities reflect the range of uncertainties in the hadronic interactions at supra-nuclear densities. This family of EOSs allows for a systematic study of the impact of stiffness on NS properties.

\subsection{Dark Matter Model}
\label{subsec:dm_model}
In this study, we employ a DM model inspired by the neutron decay anomaly, which may account for the discrepancy in the neutron lifetime measured by two distinct experimental techniques: beam experiments \cite{Czarnecki_beam-expt_2018} and bottle experiments \cite{Gonzalez_bottle-expt_2021}. According to Fornal and Grinstein \cite{Fornal_anomaly_dark_2018}, if 1\% of neutrons decay into the dark sector, this anomaly could be resolved. They proposed a decay channel where a neutron decays into a light dark boson ($\phi$) and a fermion ($\chi$) with baryon number 1:
\[
    n \longrightarrow \chi + \phi.
\]

Nuclear stability imposes constraints on the masses of these dark particles, requiring that $937.993 \, \text{MeV} < m_\chi + m_\phi < 939.565 \, \text{MeV}$ \cite{Fornal_nucl-stability_2020}. Furthermore, for $\chi$ and $\phi$ to be viable DM candidates, their stability requires the mass bound $ |m_\chi - m_\phi| < m_p + m_e = 939.783 \, \text{MeV} $ \cite{Fornal_nucl-stability_2020}. In this work, we assume the scalar boson to be massless ($m_\phi = 0$) and set the mass of the dark fermion to $m_\chi = 938.0 \, \text{MeV}$. The boson $\phi$ is assumed to escape the system with minimal interaction, establishing an equilibrium condition between the neutron and the dark fermion \cite{Motta_anomaly_2018,Husain_2022,Swarnim_2023}:
\[
    \mu_\chi = \mu_n.
\]

We take into consideration DM self-interactions by incorporating vector interactions within dark particles,
\begin{equation}
    \mathcal{L}_{DM} = -g_V\bar{\chi}\gamma^\mu \chi V_\mu - \frac{1}{4}\Phi^{\mu\nu}\Phi_{\mu\nu} + \frac{1}{2}m_V^2 V_\mu V^\mu
    \label{Ldm}
\end{equation}
where $V^\mu$ is the vector boson field and $g_V$ is the DM-vector boson coupling strength. $\Phi_{\mu\nu} = \partial_{\mu} V_\nu - \partial_\nu V_\mu$ is the field tensor and $m_V$ is the mass of the vector boson.

The energy density corresponding to the SIDM is given by, 
\begin{equation}
    {\cal E}_{DM} = \frac{1}{\pi^2} \int_{0}^{k_\chi} k^2\sqrt{k^2 + m_\chi^2} \,dk + \frac{1}{2}Gn_\chi^2
\end{equation}
where $G = \left(\frac{g_V}{m_V}\right)^2$ and $n_\chi = \frac{k_\chi^3}{3\pi^2}$ represent the self-interaction strength and number density of DM, respectively.
The chemical potentials for the DM can be derived as follows,
\begin{equation}
    \mu_\chi = \sqrt{k_\chi^2 + m_\chi^2} + Gn_\chi
\end{equation}
\begin{figure}
   \centering
   \includegraphics[width = 0.5\textwidth]{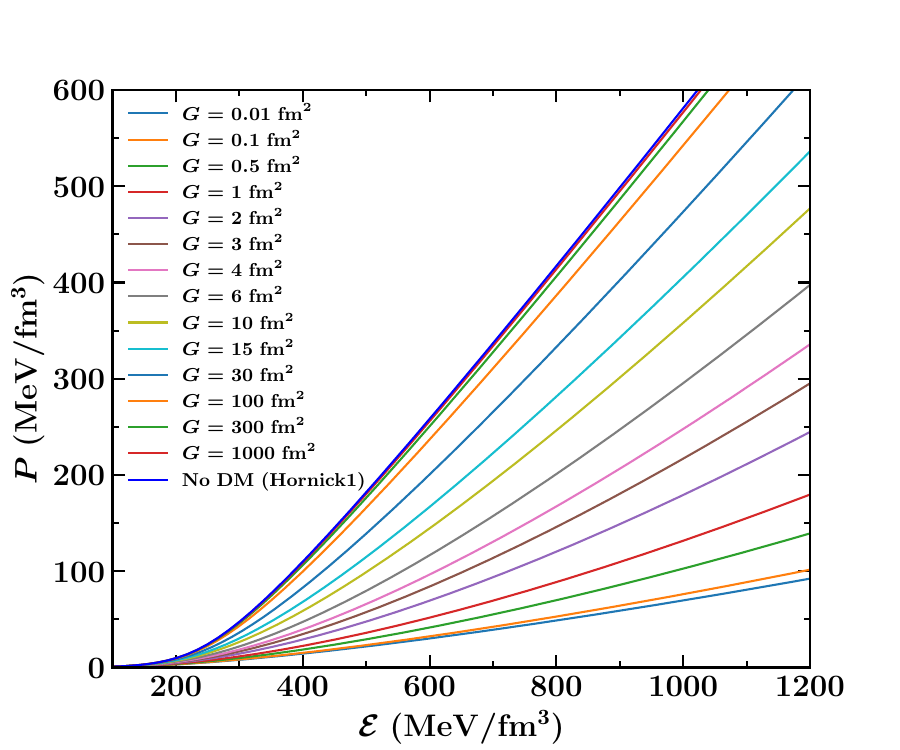}
   \caption{The EOS for DM-admixed NS is shown with varying self-interaction strength, along with without DM for Hornick1 EOS.}
    \label{fig:eos_dm_hnk1}
\end{figure}
In the Fig. \ref{fig:eos_dm_hnk1}, we considered the Hornick1 EOS to obtain the NS EOS admixed with DM. Incorporation of DM within NS is validated by varying the self-interaction strength $G$. As the parameter $G$ increases, the effect of DM on the EOS decreases. At lower values of $G$, the presence of DM significantly softens the EOS, thereby reducing the pressure for a given energy density. Conversely, as $G$ increases, the EOS gradually stiffens, and for sufficiently large $G$, it approaches the curve of pure hadronic matter (without DM). This behavior indicates that a higher DM self-interaction strength reduces the influence of DM on the NS, causing the EOS to closely resemble that of a NS devoid of DM.

\subsection{Structural Equations}
\label{subsec:steq}
\subsubsection{Static Limit}
The TOV equations, derived from Einstein's field equations in Schwarzschild-like coordinates, describe the hydrostatic equilibrium of non-rotating NSs. These equations are given by $(G = c = 1)$ \cite{Tolman_1939,Oppenheimer_1939}:
\begin{align}
    \frac{dP}{dr} &= - \frac{1}{r} \frac{({\cal E} + P)(M + 4\pi r^3 P)}{r - 2M}, \\
    \frac{dM}{dr} &= 4 \pi r^2 {\cal E} \, .
    \label{eq:tov}
\end{align}

To solve these coupled differential equations numerically, we use boundary conditions: $m(r=0) = 0$, $P(r=0) = P_c$; and $m(r=R) = M$, $P(r=R) = 0$, where $P_c$ is the central pressure and $R$ is the radius of the NS. The solution yields the mass and radius of the static NS. 

\subsubsection{Rotating Neutron Star}
\label{subsec:rns}
The conservation of angular momentum during the core collapse of a high-mass star leads to the rotation of the resulting NS. The rotation of an NS is described using the Hartle-Thorne formalism, which accounts for the effects of rotation on the spacetime metric \cite{Hartle_1967,Hartle-Thorne_1968}. For a slowly rotating star, the metric can be expressed as a perturbed Schwarzschild metric \cite{Pattersons-RNS_2021,Lopes-RNS_2024}. The metric is given by:
\begin{equation}
\label{rns-metric}
    \begin{aligned}
       ds^2 &= -e^{2\varphi}(1 + 2(h_0(r) + h_2(r)P_2(\cos\theta)))dt^2 \\
       & \quad + \bigg [1 + \frac{2(m_0(r) + m_2(r)P_2(\cos\theta))}{r - 2M(r)} \bigg ] \bigg [ 1 - \frac{2M(r)}{r} \bigg ]^{-1}dr^2 \\
       & \quad + r^2(1 + 2(k_0(r) + k_2(r)P_2(\cos\theta))) \\
       & \quad \times [d\theta^2 + \sin^2\theta(d\phi - \omega dt)^2].  
    \end{aligned}
\end{equation}

In this expression, $h_0(r)$, $h_2(r)$, $m_0(r)$, $m_2(r)$, $k_0(r)$, and $k_2(r)$ are perturbation functions that describe the deviations from the Schwarzschild metric due to rotation. $P_2(\cos\theta)$ denotes the second-order Legendre polynomial, which accounts for the angular dependence of these perturbations. The parameter $\omega$ represents the angular velocity of the local inertial frame, describing how the star's rotation affects the spacetime geometry.

Now, one may calculate the relative angular velocity $\Bar{\omega} = \Omega-\omega$ by solving the following, 
\begin{equation}
\label{omegabar}
 \frac{1}{r^4}\frac{d}{dr}\bigg ( r^4 j \frac{d\bar{\omega}}{dr} \bigg ) + \frac{4}{r}\frac{dj}{dr}\bar{\omega} = 0,
\end{equation}

using the condition
\begin{equation}
j =  j(r) = e^{-\varphi} \bigg [1 - \frac{2M(r)}{r} \bigg ]^{1/2}.
\end{equation}

Here, for the spherical star having radius $R$, we can express the angular momentum ($J$) as follows,
\begin{equation}
  J = \frac{1}{6}R^4\bigg (\frac{d\bar{\omega}}{dr} \bigg ) \bigg |_{r =R}  
\end{equation}

For the rotating star, the perturbed energy density (${\cal E}$), pressure ($P$) and number density ($n$) can be expressed as follows,
\begin{eqnarray}
\label{pertub2}
  \Delta P = ({\cal E} +P)[P_0^* + P_2^*P_2(\cos(\theta)] , \nonumber \\
  \Delta {\cal E} = ({\cal E} +P)[P_0^* + P_2^*P_2(\cos(\theta)](d{\cal E}/dP) , \nonumber \\ 
  \Delta n = ({\cal E} +P)[P_0^* + P_2^*P_2(\cos(\theta)](dn/dP).
\end{eqnarray}

Now Einstein's field equations can be solved to determine the perturbative terms by utilizing the boundary conditions at the origin, i.e., $m_0(0) = P_0^*(0) = h_2(0) = \nu_2(0) = P_2^*(0) = 0$ as follows,
\begin{equation}
\frac{dm_0}{dr} = 4\pi r^2 \frac{d{\cal E}}{dP}({\cal E} + P)P_0^* + \frac{1}{12}j^2r^4 \bigg (\frac{d \bar{\omega}}{dr} \bigg )^2  
 - \frac{1}{3} r^3 \frac{dj^2}{dr} \bar{\omega}^2, \label{dmo}
\end{equation}

\begin{equation}
\label{dp0}
    \begin{aligned}
        \frac{dP_0^*}{dr} & = - \frac{m_0(1 +8\pi r^2 P)}{(r - 2M(r))^2} - \frac{P_0^* ({\cal E} +P)4\pi r^2 }{(r -2M(r))} \\
        & + \frac{1}{12} \frac{r^4j^2}{(r - 2M(r))} \bigg ( \frac{d\bar{\omega}}{dr} \bigg )^2 + \frac{1}{3} \frac{d}{dr} \bigg ( \frac{r^3 j^2 \bar{\omega}^2}{r - 2M(r)} \bigg )
    \end{aligned}
\end{equation}

\begin{equation}
\label{dnu2}
    \begin{aligned}
        \frac{d\nu_2}{dr} & = -2h_2 \bigg ( \frac{d\varphi}{dr} \bigg ) + \bigg ( \frac{1}{r} +\frac{d\varphi}{dr} \bigg ) 
        \bigg [- \frac{1}{3} r^3 \frac{dj^2}{dr} \bar{\omega}^2   \\
        & + \frac{1}{6} j^2r^4 \bigg (\frac{d \bar{\omega}}{dr} \bigg )^2  \bigg ] ,
    \end{aligned}
\end{equation}

\begin{eqnarray}
\label{dh2}
    \begin{aligned}
        \frac{dh_2}{dr} & = -2h_2 \bigg ( \frac{d\varphi}{dr} \bigg )  + \frac{r}{r - 2M(r)} \bigg (2 \frac{d\varphi}{dr} \bigg )^{-1} \\
        & \bigg [ 8\pi ({\cal E} + P) - \frac{4M(r)}{r^3} \bigg ]h_2 - \frac{4\nu_2}{r(r -2M(r))} \bigg (2 \frac{d\varphi}{dr} \bigg )^{-1} \\
        & + \frac{1}{6} \bigg [ \frac{d\varphi}{dr}r - \frac{1}{r -2M(r)} \bigg (2 \frac{d\varphi}{dr} \bigg )^{-1}\bigg ]r^3j^2 \bigg (\frac{d \bar{\omega}}{dr} \bigg )^2 \\
        & - \frac{1}{3} \bigg [ \frac{d\varphi}{dr}r - \frac{1}{r -2M(r)} \bigg (2 \frac{d\varphi}{dr} \bigg )^{-1}\bigg ]r^3 \bigg (\frac{dj^2}{dr} \bigg ) \bar{\omega}^2 ,
    \end{aligned}
\end{eqnarray}  
\begin{equation}
 P_2^* = -h_2 - \frac{1}{3}r^2 e^{-2\varphi} \bar{\omega}^2.
\end{equation}

Now, the correction in mass can be written as,
\begin{equation}
  \delta M = m_0 + \frac{J^2}{R^3},  
\end{equation}

and also deformation can be demonstrated as follows,
\begin{equation}
 \delta r(r,\theta) =     \xi_0(r) + \xi_2(r)P_2(\cos\theta),
\end{equation}

whereas
\begin{eqnarray}
\label{xis}
  \xi_0 = -P_0^*({\cal E} + P)\bigg (\frac{dP}{dr} \bigg )^{-1}, \nonumber \\
  \xi_2 = -P_2^*({\cal E} + P)\bigg (\frac{dP}{dr} \bigg )^{-1} .
\end{eqnarray}

This allows us to determine the equator's radius ($R_e$) as well as the pole's radius ($R_p$), which is given by,
\begin{eqnarray}
\label{radii}
 R_{po} =  R + \xi_0 + \xi_2, \nonumber \\
 R_{eq} =  R + \xi_0 - \frac{1}{2}\xi_2. \nonumber 
\end{eqnarray}

Using the above two radii, we can determine the eccentricity of the star as follows,
\begin{equation}
  e =  \sqrt{ 1 - \bigg (\frac{R_{po}}{R_{eq}} \bigg )^2}. \label{ecc}  
\end{equation}

\section{Results And Discussions}
\label{sec:results}

In this section, we delve into the influence of DM on the structural properties of both the static and rotating NS.

\subsection{Static Neutron Star}
\subsubsection{Mass-Radius Relation}
\begin{figure}
\centering
   \includegraphics[width = 0.5\textwidth]{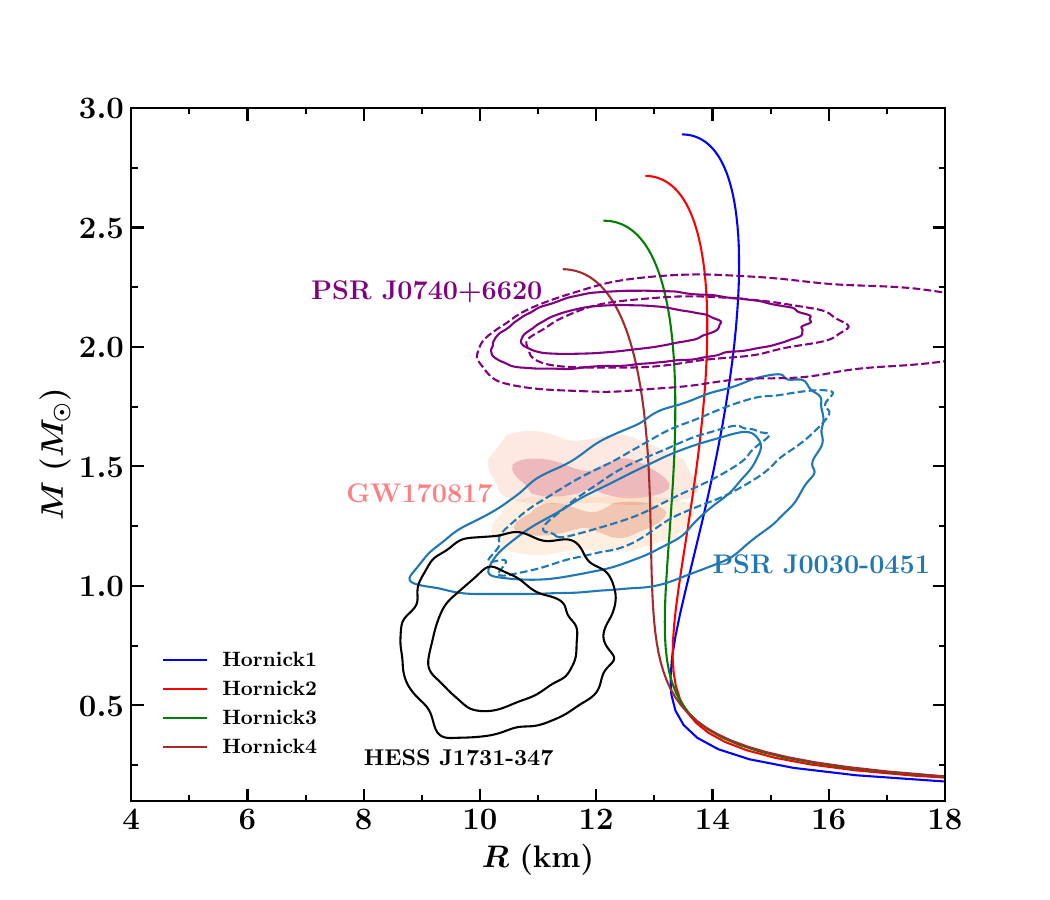}
   \caption{The mass–radius profiles for the considered EOSs are shown, along with multiple observational constraints. The $1\sigma$ and $2\sigma$ confidence contours from the NICER measurements of PSR J0030+0451 \cite{Miller_2019, Riley_2019} and PSR J0740+6620 \cite{Miller_2021, Riley_2021} are included, where the dashed and solid lines correspond to the analyses of Miller et al. and Riley et al., respectively. Also, the mass–radius constraint from the compact object in HESS J1731–347 \cite{HESS_2022} has been imposed. Additionally, the $50\%$ and $90\%$ confidence intervals obtained from the LIGO–Virgo analysis of the binary neutron star merger event GW170817 \cite{Abbott_2018} are also used.}
    \label{fig:mr_hornick}
\end{figure}

In Fig.\ref{fig:mr_hornick}, the mass-radius profiles for static NSs are shown, considering Hornick hadronic EOSs with mass-radius constraints from PSR J0030-0451 \cite{Miller_2019,Riley_2019}, PSR J0740+6620 \cite{Miller_2021,Riley_2021}, HESS J1731-347 \cite{HESS_2022} and GW170817 \cite{Abbott_2018}. All EOSs exceed PSR J0740+6620's upper mass limit due to their stiffness at high densities. To satisfy HESS J1731-347 along with $2 \ M_\odot$ constraint, the EOS model must be sufficiently soft at low density and stiff at high density. To ensure compatibility with pulsar mass-radius constraints, contributions like hyperons or phase transitions are often needed. Here, we consider DM as an additional element to help reduce NS maximum mass, aligning DM-admixed NSs with astrophysical constraints.

\begin{figure}
\centering
   \includegraphics[width = 0.5\textwidth]{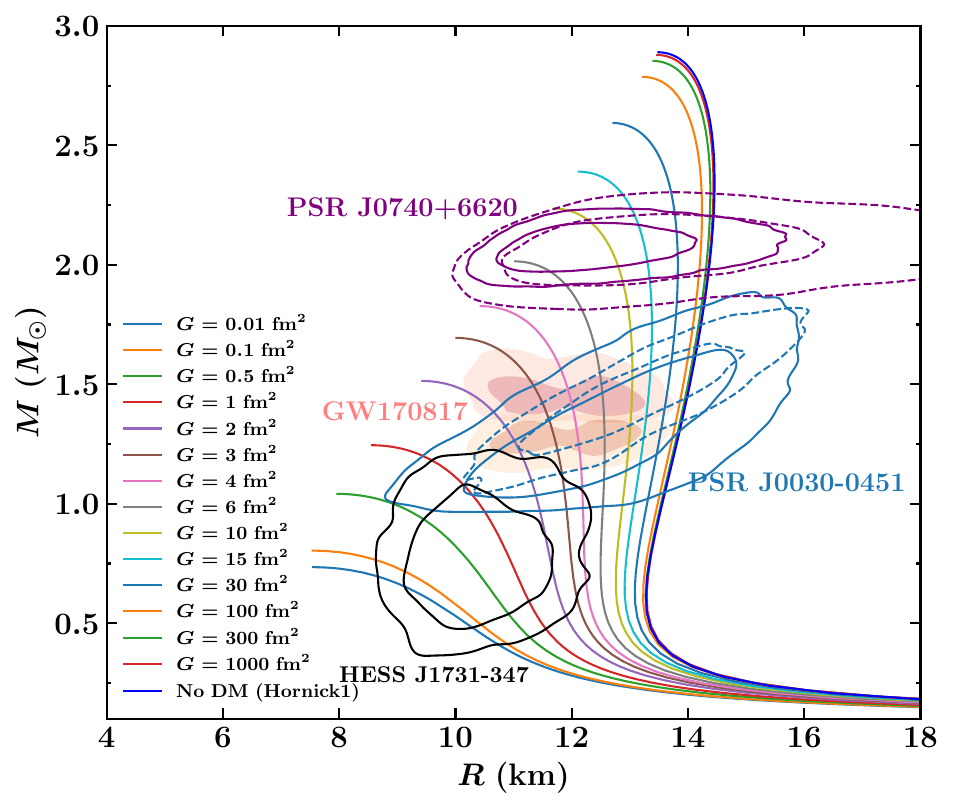}
   \caption{The profile of mass-radius with different strengths of self-interaction is displayed and contrasted with the situation without DM for Hornick1 EOS. Similar to Fig. \ref{fig:mr_hornick}, the mass-radius constraints from PSR J0030+0451 \cite{Miller_2019, Riley_2019}, PSR J0740+6620 \cite{Miller_2021, Riley_2021}, HESS J1731–347 \cite{HESS_2022} and GW170817 \cite{Abbott_2018} are imposed to constrain our model predictions.}
    \label{fig:mr_hnk1}
\end{figure}
To study DM impact on the mass-radius profile, we chose the Hornick1 EOS and treated the DM interaction strength $G$ as a free parameter, using the neutron decay anomaly model from Sec. \ref{subsec:dm_model}. We compared the mass-radius profiles of pure hadronic stars with and without DM (``No DM''). The maximum mass and corresponding radius, and also the canonical radius reduced effectively with the inclusion of the DM. To extract the viable models of DM-admixed NS, and to maintain the consistency between the DM-admixed NS results and astrophysical data, we have imposed various observational constraints for mass-radius profile such as PSR J0030-0451 \cite{Miller_2019, Riley_2019}, PSR J0740+6620 \cite{Miller_2021, Riley_2021}, HESS J1731-347 \cite{HESS_2022} and GW170817 \cite{Abbott_2018}. As it is observed that in the Fig. \ref{fig:eos_dm_hnk1}, the effect of DM is efficient for lower interaction strengths and diminishes for higher interaction strengths. A similar behavior was also observed in the mass-radius profile. For a certain range of interaction strengths, the mass-radius profile satisfies the constraints from PSR J0030-0451 for canonical models and the maximum mass models constraint from PSR J0740+6620, along with the GW constraint. With decreasing the value of interaction strength, the mass-radius profile satisfies the HESS J1731-347, but the maximum mass is reduced below $2 \ M_\odot$ constraint. To align with pulsar mass constraints, a strict limit on the interaction strength is necessary.

\subsubsection{Maximum Mass And DM fraction}
\begin{figure}
\centering
   \includegraphics[width = 0.5\textwidth]{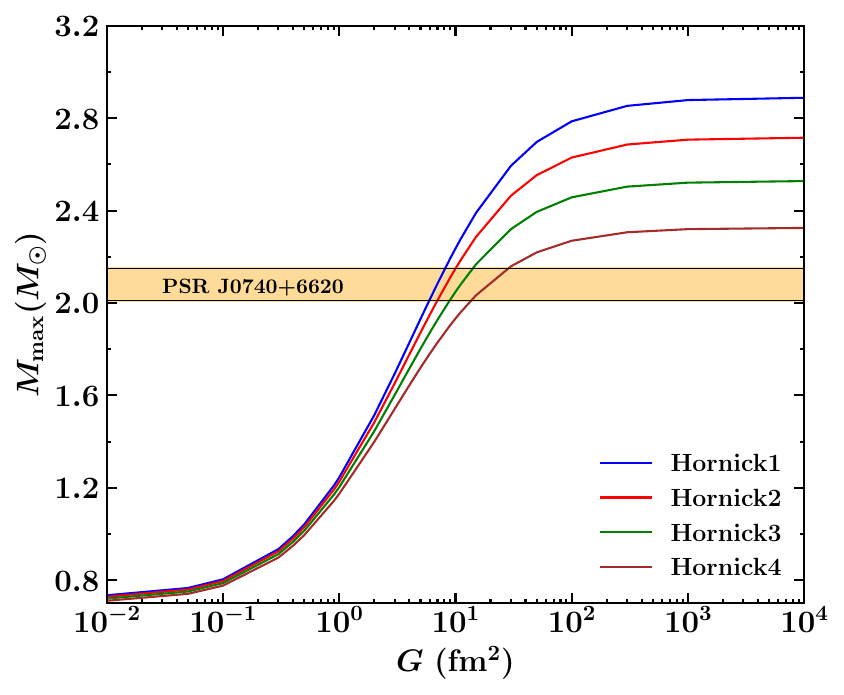}
   \caption{The maximum mass of considered EOSs shown with the variation of interaction strength $G$. The orange shaded bar denotes the mass limit of pulsar PSR J0740+6620 \cite{Miller_2021}.}
    \label{fig:MG_hornick}
\end{figure}

In Fig. \ref{fig:MG_hornick}, the maximum mass of a DM-admixed NS is plotted as a function of interaction strength for four considered EOSs. The maximum mass of each EOS is highly reduced for lower interaction strengths, while in increasing the interaction strengths increases the maximum mass. The maximum mass of each EOS falls within the maximum mass constraint of different pulsars with a certain range of interaction strengths. In this study, we determine the range of interaction strengths for each EOS utilizing likelihood analysis by imposing various astrophysical constraints, which will be discussed in the upcoming section.

\begin{figure}
\centering
   \includegraphics[width = 0.5\textwidth]{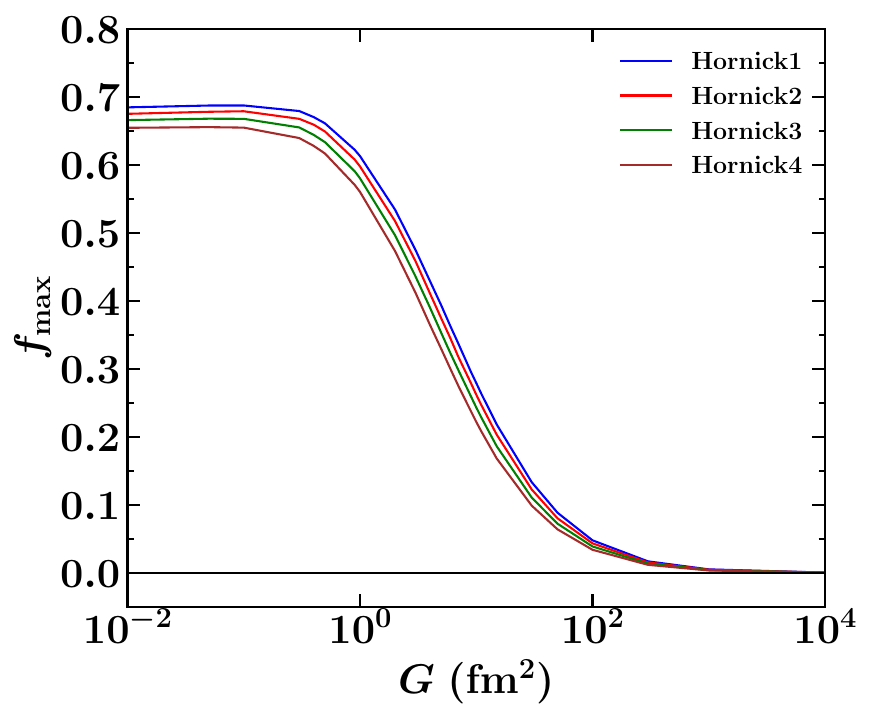}
   \caption{The maximum DM mass fraction is plotted with DM interaction strength.}
    \label{fig:fx_G_Hornick}
\end{figure}

As observed in Fig. \ref{fig:mr_hnk1}, the effect of DM diminishes with increasing interaction strength ($G$). Therefore, it is essential to calculate the DM fraction $(f_{\rm DM} = M_{\rm DM}/M = \int_{0}^{R} 4\pi r^2 \varepsilon_{\rm DM} dr/\int_{0}^{R} 4\pi r^2 \varepsilon dr)$, which quantifies the fraction of DM in a DM-admixed NS. In the upper panel of Fig. \ref{fig:fx_G_Hornick}, the DM fraction is plotted as a function of $G$ for the maximum mass configuration $(f_{\rm max})$. For each EOS, it is observed that as the DM interaction strength increases, the corresponding DM fraction decreases, approaching nearly $0\%$ at sufficiently high $G$. One can estimate the corresponding maximum DM fraction for each EOS after the calculation of the range of DM interaction strength using observational constraints.

\subsubsection{Likelihood Analysis}

Here, we have conducted a likelihood analysis to constrain the DM interaction strength for DM-admixed NS EOS. This analysis was carried out by utilizing the astrophysical constraints from NICER X-ray observations of the pulsars PSR J0030+0451 \cite{Miller_2019, Riley_2019} and PSR J0740+6620 ~\cite{Miller_2021, Riley_2021}. The Gaussian Kernel density estimator (KDE) has been employed to model the probability distribution of DM-admixed NS properties.

Numerous studies have recently employed this method to build likelihoods for investigating the astrophysical data from pulsar and GW observations \cite{Raaijmakers:2019dks, Biswas:2020puz, Vinciguerra:2023qxq, Imam:2024gfh}. The X-ray observations provide measurements of the NS mass ($M$) and radius ($R$), enabling the likelihood to be formulated as:

\begin{align}
    P(d_{\rm X-ray}|\text{EOS}) = \int^{M_{\rm max}(\text{EOS})} dm \int dR \, \delta(R - R(\text{EOS}, m)) \nonumber \\
    \times P(d_{\rm X-ray} | m, R). \label{eq:likelihood}
\end{align}

Here, $ M_{\rm max}(\text{EOS}) $ represents the maximum mass of a NS for a given EOS and DM interaction strength $G$. The likelihood function in Eq.(\ref{eq:likelihood}) quantifies the probability of observing the NICER X-ray data given a specific EOS. By integrating over possible NS masses up to $ M_{\rm max}(\text{EOS}) $, it ensures that only physically viable solutions are considered, while the Dirac delta function links the mass-radius relationship inherent to the EOS. KDE is employed to smooth out the likelihood data, offering a non-parametric approach to estimate the probability distribution of NS properties, which helps in constraining the DM interaction strength $G$ within a DM-admixed EOS.

\begin{figure}
    \centering
    \includegraphics[width = 0.5\textwidth]{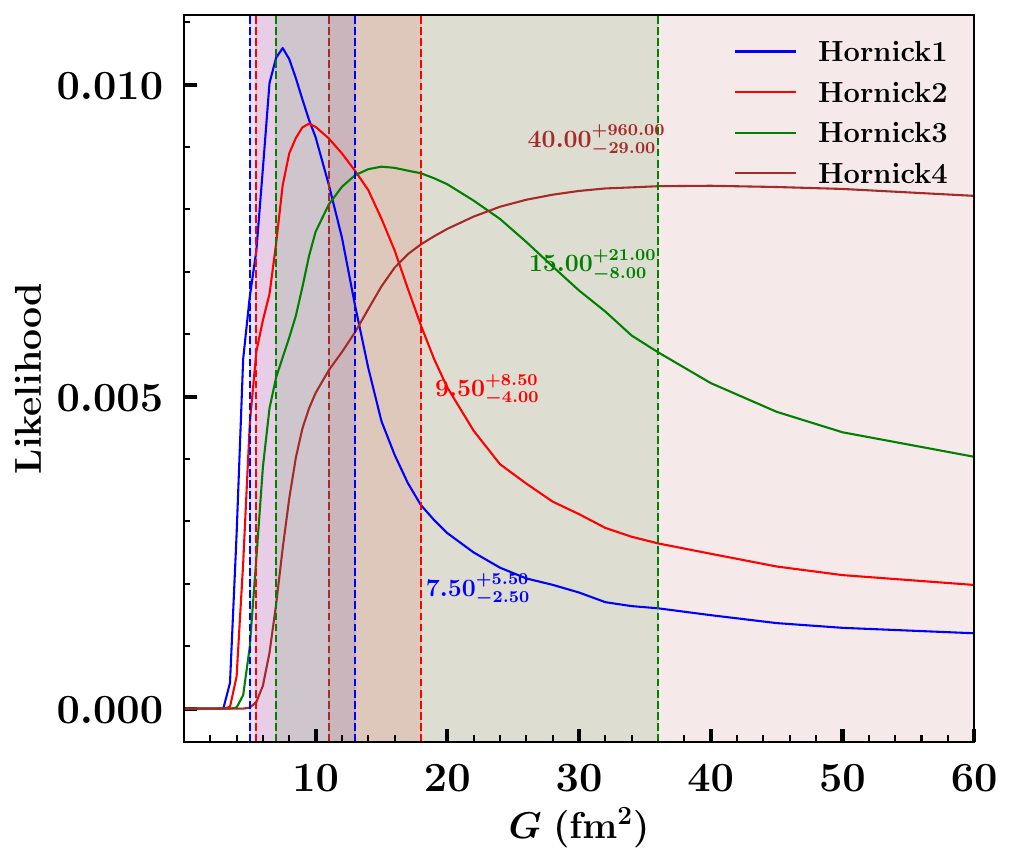}
    \caption{Likelihood probability distribution of DM EOSs as a function of the DM interaction strength ($G$). The shaded regions represent the $68\%$ confidence interval (CI), with the corresponding range of $G$ values.}
    \label{fig:likelihood_G}
\end{figure}

Figure-\ref{fig:likelihood_G} shows the likelihood distribution as a function of DM interaction strength ($G$) for four different EOS models. The vertical lines with shaded regions represent the 68\% confidence intervals over the valid range of ($G$) values for each respective DM-admixed model. As we move from Hornick1 to Hornick4, the likelihood peak and the range of the shaded region decrease. This trend is due to the decreasing stiffness of the EOS from Hornick1 to Hornick4. A stiffer EOS, such as Hornick1, predicts larger NS radii for a given mass, which is more sensitive to the DM interaction strength, while softer EOS models like Hornick4 predict smaller radii, resulting in a broader likelihood distribution and larger uncertainties in ($G$).
The DM-admixed models with stiffer EOSs (like Hornick1) show a stronger influence of the DM interaction, while for softer EOSs (Hornick4), the likelihood function becomes less sensitive to ($G$), as observed in the wider confidence intervals. This behavior illustrates the interplay between the stiffness of the EOS and the required DM interaction strength to match the NICER data. Specifically, the Hornick1 EOS favors a lower ($G$) value of $7.50^{+5.50}_{-2.50} \, \text{fm}^2$, while the Hornick4 EOS suggests a higher ($G$) value of $40.00^{+960.00}_{-29.00} \, \text{fm}^2$. These results highlight the sensitivity of the likelihood function to the choice of EOS, underlining the critical role of EOS stiffness in determining the DM interaction strength that satisfies the observed NS properties. The constrained ($G$) values with 68\% confidence intervals are also listed in Table-\ref{tab:G-values}, providing a detailed view of the dependence of DM interaction strength on the EOS models.

\begin{figure}
    \centering
    \includegraphics[width = 0.5\textwidth]{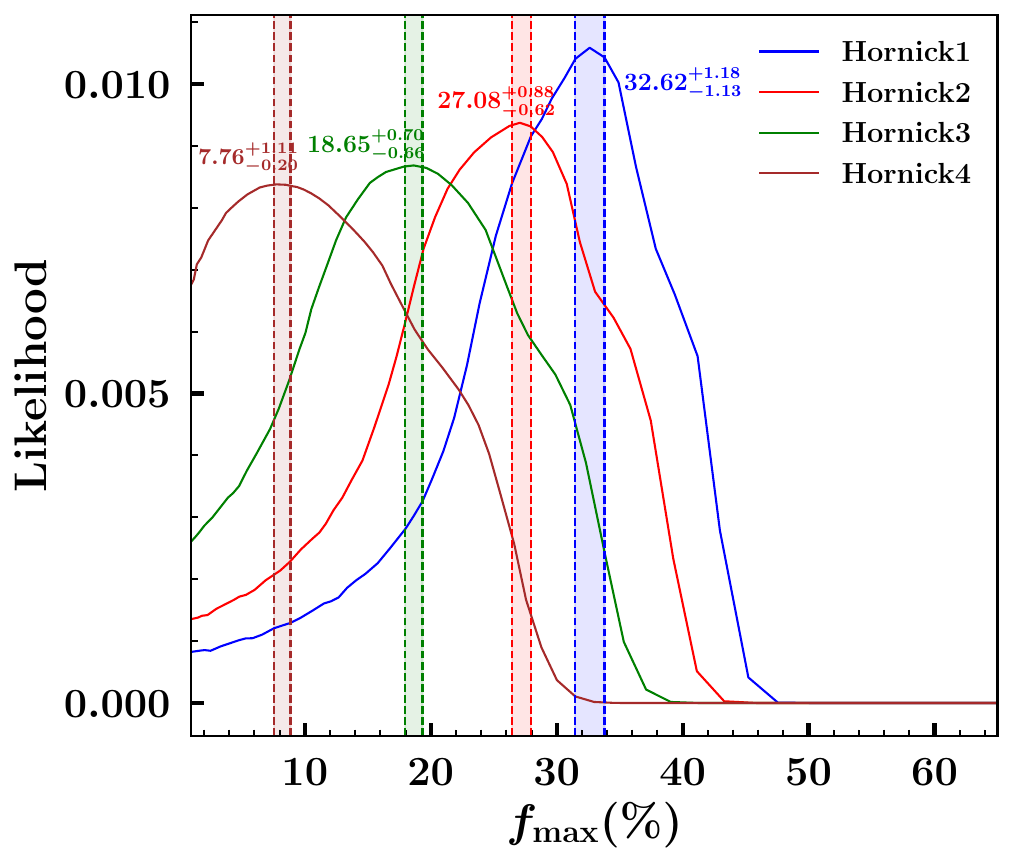}
    \caption{The likelihood probability distribution for DM EOSs is shown as a function of the maximum mass fraction. Shaded regions denote the $68\%$ confidence interval, along with the corresponding range of $f_{\rm max}$ values.}
    \label{fig:likelihood_fx}
\end{figure}

Similar to Fig. \ref{fig:likelihood_G}, we have constrained the maximum DM mass fraction for each model using likelihood analysis, as shown in Fig. \ref{fig:likelihood_fx}. The $68\%$ confidence interval is highlighted to indicate the allowed range of $f_{\rm max}$. Our results show that the Hornick1 model retains the highest fraction, while the Hornick4 model retains the lowest among all four models. The fraction decreases with decreasing stiffness of the model, which is in contrast to the behavior observed for the interaction strength. The constrained values with their corresponding $68\%$ confidence intervals are summarized in Table \ref{tab:G-values}, giving a clearer picture of how the maximum DM mass fraction depends on the stiffness of the EOS model.

\begin{table}
\centering
\caption{Preferred values of the DM interaction strength ($G$) and DM maximum mass fraction ($f_{\rm max}$) with 68\% confidence intervals for the EOS models analyzed in this study.}
\renewcommand{\tabcolsep}{0.3cm}
\renewcommand{\arraystretch}{1.2}
\scalebox{1.20}{
\begin{tabular}{ccccc}
\hline \hline
 & Hornick1 & Hornick2 &  Hornick3   & Hornick4   \\
\hline
 G ($ \text{fm}^2 $) & $7.50^{+5.50}_{-2.50}$ & $9.50^{+8.50}_{-4.00}$ & $15.0^{+21.0}_{-8.0}$ & $40.0^{+960.0}_{-29.0}$ \\
 $f_{\rm max}$ ($\%$) & $32.62^{+1.18}_{-1.13}$ & $27.08^{+0.88}_{-0.62}$ & $18.65^{+0.70}_{-0.66}$ & $7.76^{+1.11}_{-0.20}$ \\
\hline \hline
\end{tabular}}
\label{tab:G-values}
\end{table}

\subsection{Rotating Neutron Star}

Before interpreting the results of the rotating NS analysis, it is crucial first to understand how the properties of non-rotating NSs relate to the Keplerian frequency. This understanding is important because determining the Keplerian frequency is essential for identifying the maximum rotational speed that a NS can sustain before becoming unstable. The Keplerian frequency serves as a critical parameter, defining the threshold of angular velocity—the mass-shedding limit—beyond which the star would begin to lose mass. For a relativistic star, the Keplerian frequency is expressed as follows \cite{Glendenning_rns_1992, Lopes-RNS_2024}:
\begin{equation}
    \Omega_k = 0.65 \sqrt{\frac{M}{R^3}}
\end{equation}

\begin{figure}
   \centering
   \includegraphics[width = 0.5\textwidth]{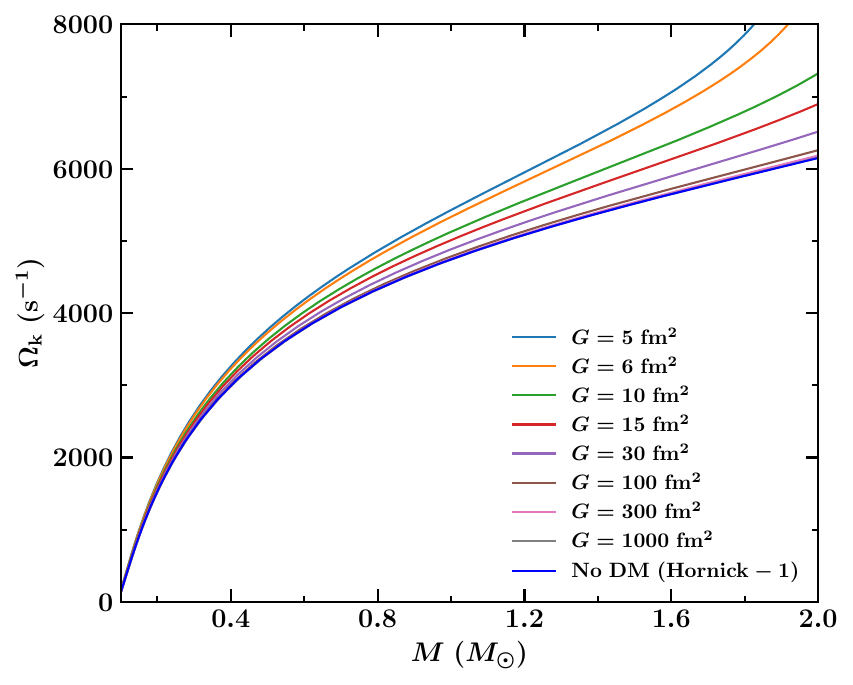}
   \caption{The Keplerian frequency is shown as a function of static mass of the NS.}
    \label{fig:kepler-mass}
\end{figure}

Fig. \ref{fig:kepler-mass} shows how the Keplerian frequency ($\Omega_k$) relates to the mass of a static NS for the Hornick1 EOS, highlighting DM effects with different interaction strengths $G$. A key observation is that as the angular velocity increases, the mass-shedding limit also increases, meaning that a faster-rotating star can sustain a higher mass before shedding begins. Regarding the DM effects, DM-admixed NS allows higher Keplerian frequencies at the same mass compared to those without DM. Higher DM interaction strengths raise the mass-shedding limit at constant frequency, while lower strengths reduce it. Keeping the limit of angular velocity in light of the observed fastest pulsar PSR J1748-2446ad ($\Omega = 4498 \ {\rm s^{-1}}$), we choose two angular velocities for RNS analysis: $\Omega = 3000 \ {\rm s}^{-1}$ and $4500 \ {\rm s}^{-1}$.

\begin{figure}
   \centering
   \includegraphics[width = 0.5\textwidth]{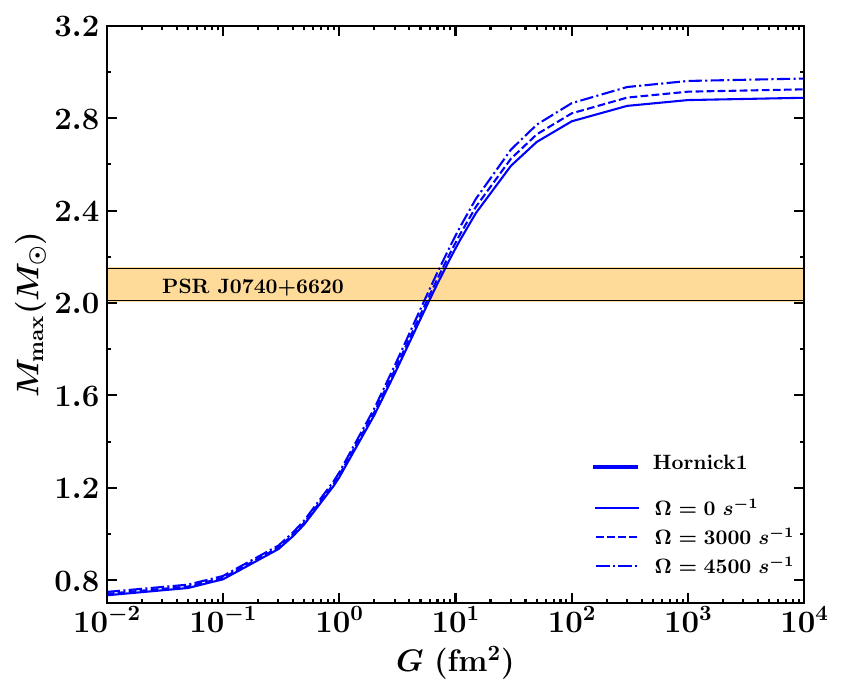}
   \caption{The maximum mass as a function of DM interaction strength is shown at different angular velocities.}
    \label{fig:M_G_hnk1}
\end{figure}

To understand the influence of DM on the maximum mass of rotating NS, we plotted the maximum rotational mass against DM interaction strength at various angular velocities in Fig. \ref{fig:M_G_hnk1}. The static case ($\Omega=0$) is included for comparison. As rotation begins, the maximum mass of the DM-admixed NS increases, especially at higher DM interaction strengths. As the maximum mass of the NS rises with the rotation, the constraints on $G$ will be change to meet the observation constraints. As the rotational deformation affects the NS structure, it will modify the likelihood profiles of DM DM-admixed rotating NS. Hence, the likelihood analysis also required investigating the DM-admixed rotating NS.

\begin{figure}
    \centering
    \includegraphics[width=0.5\textwidth]{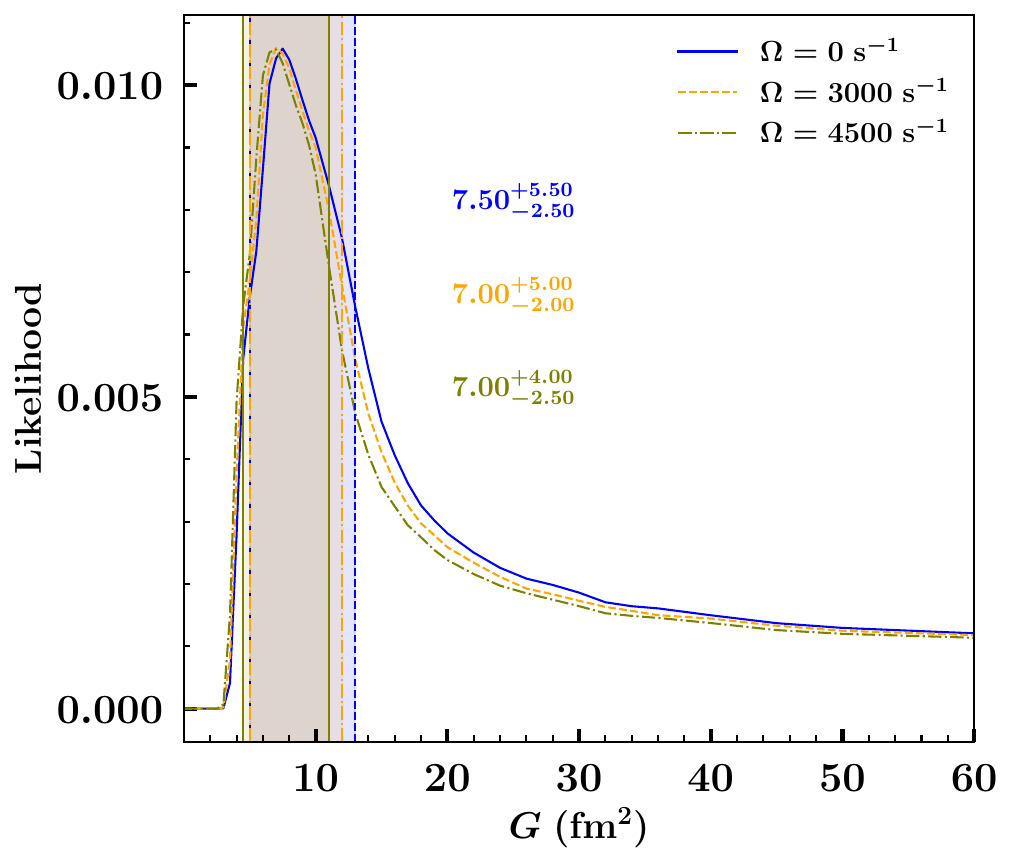}
    \includegraphics[width=0.5\textwidth]{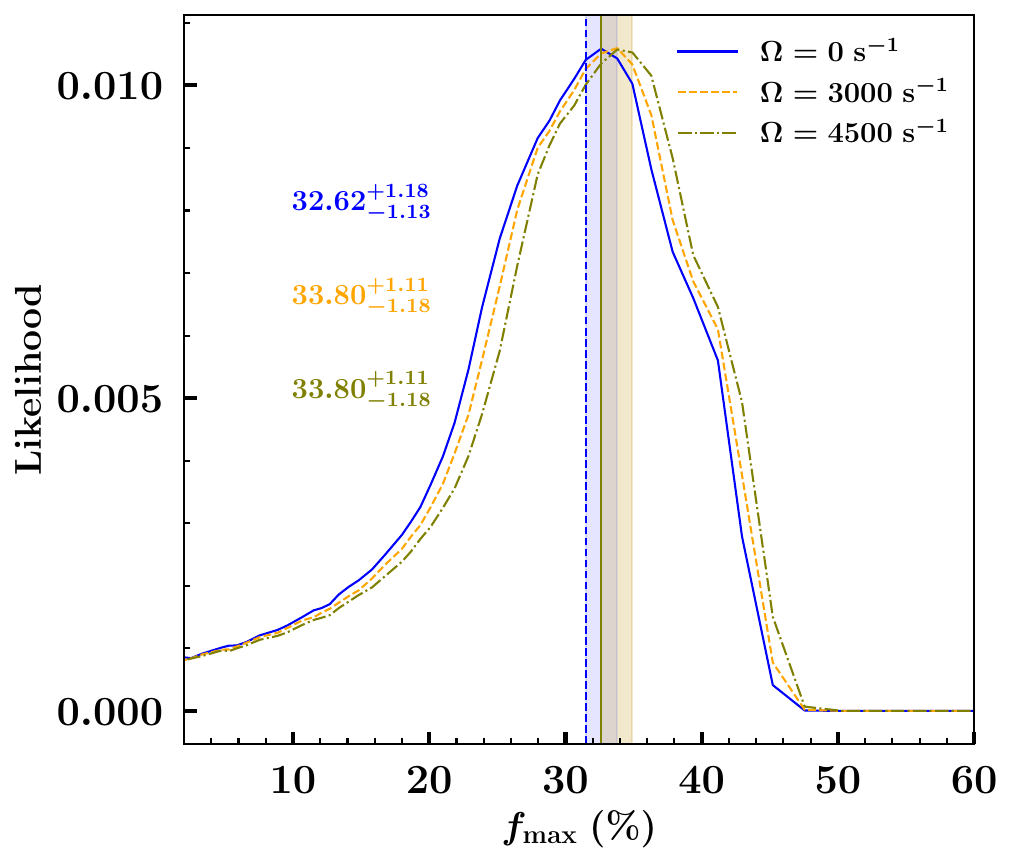}
    \caption{The likelihood probability distributions for the Hornick1 EOS at different angular velocities are shown, with the $68\%$ confidence intervals indicated, as a function of the interaction strength ({\it Left panel}) and the maximum DM mass fraction ({\it Right panel}).}
    \label{fig:likelihood_Gw}
\end{figure}

In the left panel of Fig. \ref{fig:likelihood_Gw}, the likelihood distribution for different angular velocities of the NS: $\Omega = 0\, \text{s}^{-1}$, $\Omega = 3000\, \text{s}^{-1}$, and $\Omega = 4500\, \text{s}^{-1}$ shown. As the maximum mass increases with angular velocity, the range of DM interaction reduces to meet the observational constraints. For static case, the value of DM interaction was, $G = 7.50^{+5.50}_{-2.50}\, \ \text{fm}^2$, while for  $\Omega = 3000\, \text{s}^{-1}$ and  $\Omega = 4500\, \text{s}^{-1}$, it reduces to $7.00^{+5.00}_{-2.00}\, \text{fm}^2$ and $7.00^{+4.00}_{-2.50}\, \text{fm}^2$, respectively. Similarly, in the right panel, at different angular velocities, we perform the likelihood analysis as a function of maximum DM mass fraction. The results show that the increase in angular velocity increases the $f_{\rm max}$ as compared to the static case. The static case favors $f_{\rm max} = 32.62^{+1.18}_{-1.13}$, while both $\Omega = 3000,\mathrm{s^{-1}}$ and $\Omega = 4500,\mathrm{s^{-1}}$ peak at essentially the same value, $f_{\rm max} = 33.80^{+1.11}_{-1.18}$. This indicates that rotating neutron stars are capable of retaining a higher fraction of DM than non-rotating ones.

\begin{figure}
   \centering
   \includegraphics[width = 0.5\textwidth]{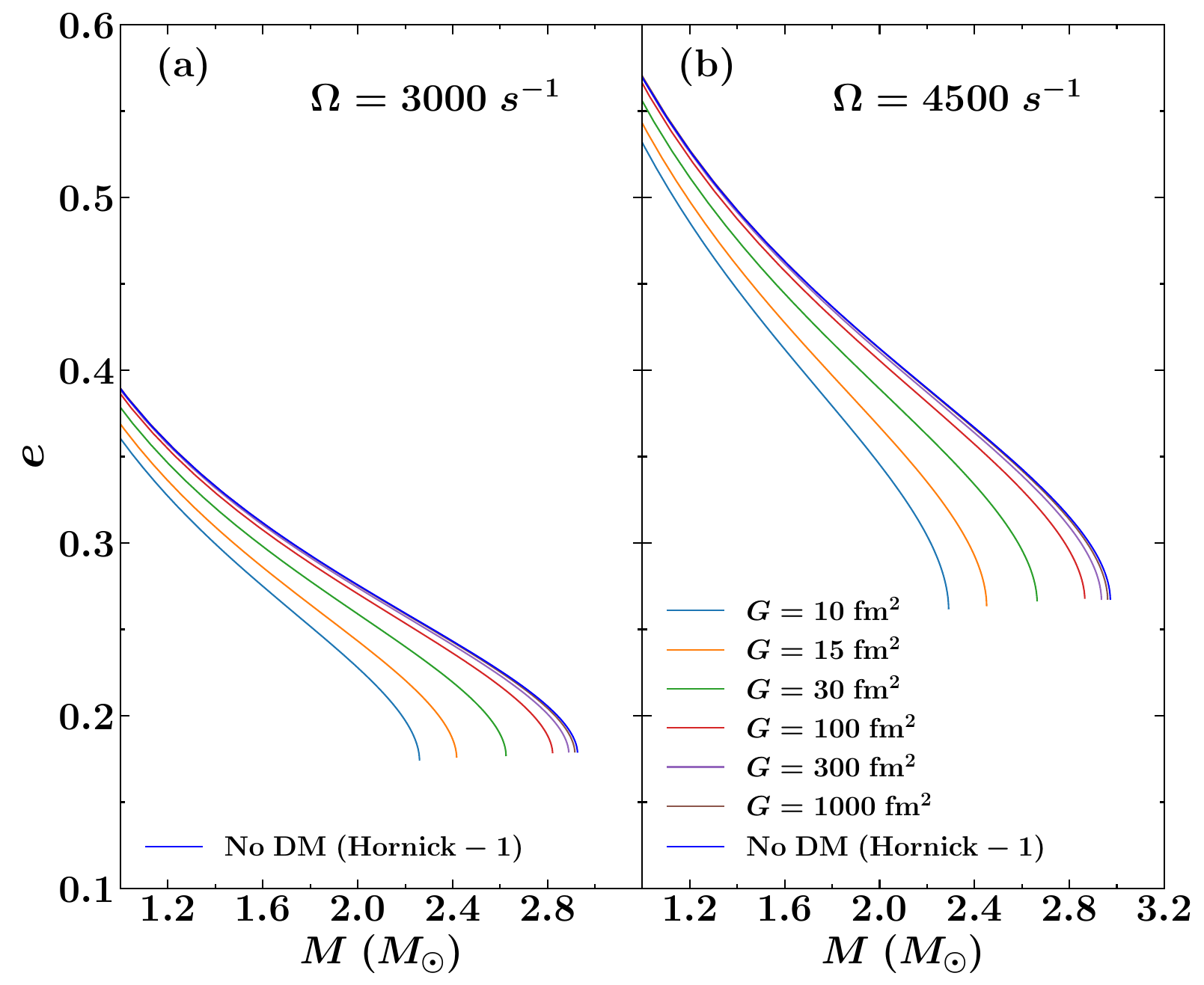}
   \caption{For different rotating limits, the eccentricity is shown as a function of mass.}
    \label{fig:eccen-m}
\end{figure}

Figure \ref{fig:eccen-m} illustrates the variation of eccentricity ($e$) with rotational mass. Eccentricity is a key parameter to study the deformation of the NS, which relates both polar and equatorial radius. For a static star ($\Omega = 0 \ s^{-1}$), the NS is considered spherical denotes that both polar and equatorial radii are the same, and hence $e=0$. When the star starts to rotate with higher angular velocity, as shown in panel-a ($\Omega = 3000 \ s^{-1}$) and panel-b ($\Omega = 4500 \ s^{-1}$), the NS deviates from its spherical shape with a finite value of eccentricity. Higher eccentricity associated with higher angular velocity denotes significant deviations from spherical shape, leading to more oblate stars. The effects of the DM on the deformation were also observed with varying $G$. A lower value of $G$ reduces the deformation of rotating NS as compared to higher values. Hence, it can be concluded that a rotating DM-admixed NS shows a smaller deformation as compared to a rotating NS without DM.

\begin{figure}
   \centering
   \includegraphics[width = 0.5\textwidth]{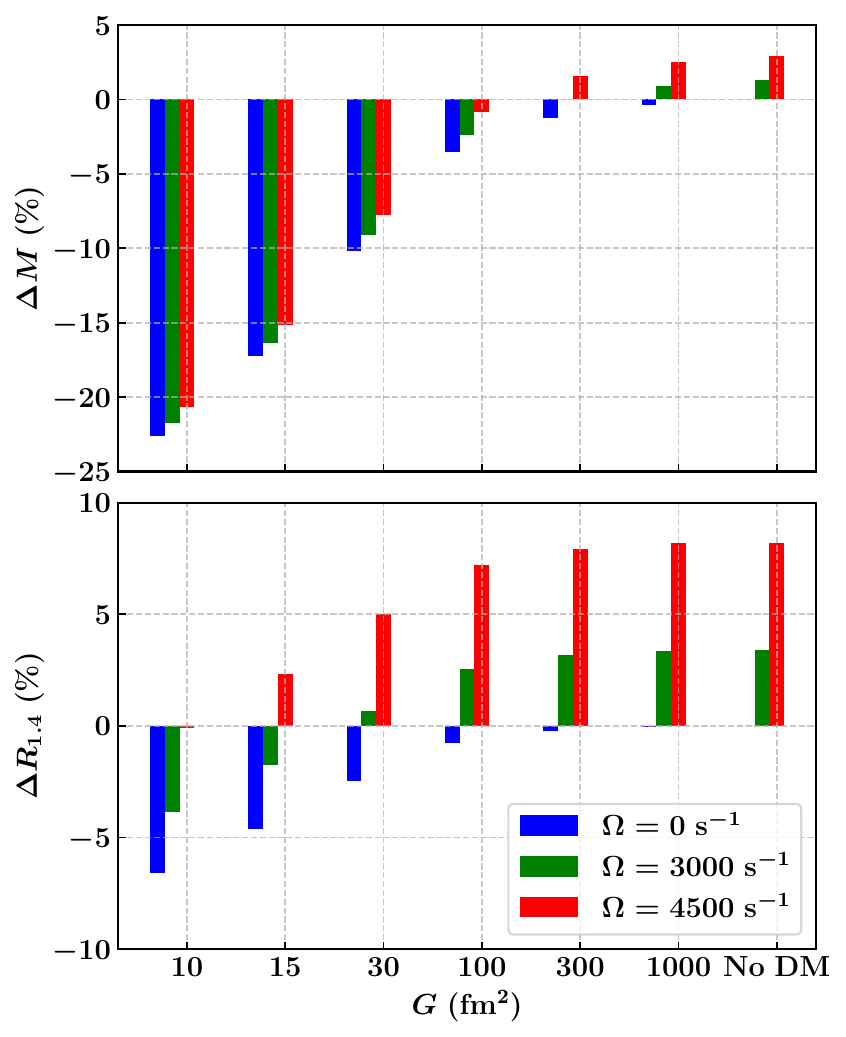}
   \caption{The relative deviation for maximum mass ($\Delta M$) and canonical radius ($\Delta R_{1.4}$) is shown.}
    \label{fig:relative-deviation}
\end{figure}

Figure \ref{fig:relative-deviation} illustrates how maximum mass ($\Delta M$) and canonical radius ($\Delta R_{1.4}$) deviate relative to the DM self-interaction strength, $G$, at three angular velocities: $\Omega = 0 \, \text{s}^{-1}$, $3000 \, \text{s}^{-1}$, and $4500 \, \text{s}^{-1}$ from their baseline values (static NS without DM). In the top panel, $\Delta M$ for the non-rotating case ($\Omega = 0 \, \text{s}^{-1}$) drops significantly, reaching 20\% below the baseline at low $G$. With increasing $G$, the deviation lessens, turning positive beyond $G = 100 \, \text{fm}^2$. Rotation reduces these deviations; at $\Omega = 3000 \, \text{s}^{-1}$ and $\Omega = 4500 \, \text{s}^{-1}$, the drop in $\Delta M$ lessens. For $\Omega = 4500 \, \text{s}^{-1}$, the maximum mass surpasses the baseline for $G \geq 100 \, \text{fm}^2$, showing rotation counteracts DM softening effect, resulting in a net positive deviation. The bottom panel presents $\Delta R_{1.4}$, showing trends similar to $\Delta M$. For $\Omega = 0 \, \text{s}^{-1}$, $\Delta R_{1.4}$ initially drops, peaking at low $G$. As angular velocity rises, this drop reduces, and for $G \geq 100 \, \text{fm}^2$, $\Delta R_{1.4}$ turns positive, indicating larger radii for rotating DM-admixed NSs. This shows that rotation enhances the equatorial radius, offsetting DM-induced compaction and causing increased deformation at higher angular velocities.

\section{Summary And Conclusions}
\label{sec:summary}
In this study, we investigate the impacts of DM on the static and rotating NS properties. A self-interacting DM model motivated by the neutron decay anomaly was employed, treating the DM interaction strength ($G$) as a variable. Four hadronic models, Hornick1, Hornick2, Hornick3, and Hornick4, are chosen, and those meet $\chi$-EFT constraints in the low-density regime and are also sufficiently stiff in the high-density regime to meet the maximum mass limits of massive pulsar PSR J0740-6620. The EOSs for DM-admixed NS are derived by varying $G$; at lower $G$, the EOS becomes significantly softer, while increasing $G$ results in a stiffer EOS. A similar behavior is observed in the mass-radius relation: lower $G$ effectively reduces the maximum mass of the DM-admixed NS, whereas higher $G$ increases it, approaching a purely hadronic case at sufficiently high $G$. Further, we constrained both the DM interaction strength and the maximum DM mass fraction using likelihood analysis informed by NICER’s mass–radius measurements of PSR J0740+6620 and PSR J0030+0451. The likelihood function was constructed across different values of $G$ and $f_{\rm max}$ using a kernel density estimator to evaluate the corresponding neutron star properties. Our results show that stiffer EOSs are more sensitive to variations in $G$, leading to narrower confidence intervals, whereas softer EOSs yield broader likelihood profiles with larger uncertainties in $G$. In contrast, the trend is reversed for $f_{\rm max}$, where softer EOSs allow tighter constraints, while stiffer ones exhibit broader ranges.

Coming to the rotating NS, the Hartle-Thorne formalism has been employed to study its properties, treating the angular velocity ($\Omega$) as a free parameter. The maximum mass of the NS increases with angular velocity. DM exhibits a similar influence on rotating NS properties as it does in the static case. Since both rotation and DM affect the NS's behavior, the bounds on the interaction strength are also impacted. The likelihood analysis was also performed for the rotational case to constrain both $G$ and $f_{\rm max}$, and the results were compared with the static case. Due to the increase in maximum mass of the NS with rotation, the allowed range of $G$ becomes narrower relative to the static case. In contrast, the opposite behavior is observed for $f_{\rm max}$, where rotation permits a broader range of values. This implies that rapidly rotating NS are capable of retaining a larger fraction of DM compared to their static case. Another key observation in the rotating NS case is the deformation of the star as it begins to rotate, deviating from its spherical shape, which is reflected in the finite value of eccentricity. DM also plays a significant role in influencing this deformation. An increase in the DM fraction within the NS reduces deformation, helping the star to retain a shape closer to spherical. So, here we can conclude that a DM-admixed rotating NS is less deformed as compared to the rotating NS without containing DM. From all the above analyses, it is remarkable that if in the future such a DM-admixed NS is observed, then it will be challenging to extract the information from this complex structure. 

\section*{Acknowledgements}
B.K. acknowledges partial support from the Department of Science and Technology, Government of India with grant no. CRG/2021/000101. NKP acknowledge support by the CUHK-Shenzhen University development fund under grant No. UDF01003041 and UDF03003041, and Shenzhen Peacock fund under No. 2023TC0007.

\bibliographystyle{apsrev4-2}
\bibliography{main}
\end{document}